\documentclass[twocolumn]{aastex701}

\usepackage{bm,enumitem} 
\usepackage{newtxmath,newtxtext,subcaption}
\usepackage{empheq}

\newcommand{\pCNOP}{{\tt CNOP}}     % CNOP optimal perturbation
\newcommand{\pRand}{{\tt Random}}   % random Gaussian perturbation
\newcommand{\pMmode}{{\tt 5modes}}  % five dominant Fourier modes

\begin{document}

\title{The Most Dangerous Seed: Nonlinear Optimal Perturbations in Rayleigh-Taylor Instability}

\author[0000-0001-9658-0588]{Suoqing Ji}
\affiliation{Center for Astronomy and Astrophysics and Department of Physics, Fudan University, Shanghai 200438, P.R.China.}
\affiliation{Key Laboratory of Nuclear Physics and Ion-Beam Application (MOE), Fudan University, Shanghai 200433, P.R.China.}
\email[show]{sqji@fudan.edu.cn}
   
\author[0000-0001-8616-6180]{Bin Shi} 
\affiliation{Center for Mathematics and Interdisciplinary Sciences, Fudan University, Shanghai 200433, P.R.China.}
\affiliation{Shanghai Institute for Mathematics and Interdisciplinary Sciences, Shanghai 200433, P.R.China.}
\email[show]{binshi@fudan.edu.cn}

\begin{abstract}
Long-term instabilities in astrophysical fluids are inherently nonlinear, where even small-amplitude perturbations can trigger dramatic instability. However, owing to the complex interactions among non-normal modes, the perturbation structures responsible for the greatest growth remain poorly understood. In this paper, we employ the nonlinear optimization method known as the conditional nonlinear optimal perturbation (CNOP) to identify the most dangerous initial velocity perturbation, i.e., the perturbation that maximizes the kinetic energy growth in the two-dimensional compressible Rayleigh-Taylor instability in astrophysical hydrodynamics. Compared with random perturbations, the optimal perturbation forms a coherent wave-packet structure localized around the density interface. We investigate its dependence on spatial resolution and optimization time horizon through two sets of numerical experiments. For a fixed optimization time, increasing the spatial resolution produces a more sharply localized wave packet, whereas for a fixed spatial resolution, increasing the optimization time causes the wave packet to become progressively more dispersed. Furthermore, we analyze the optimal perturbations in Fourier space using the fast Fourier transform (FFT), which provides a clearer characterization of the spectral distribution.  Higher-resolution simulations concentrate most of the perturbation energy into only a few dominant modes, while longer optimization times distribute the energy over a broader range of modes. These results indicate that short optimization time horizons involve relatively weak modal interactions and remain closer to the linear regime, whereas longer optimization times enhance nonlinear modal interactions, broaden the spectral distribution, and reduce the predictive capability of linear stability theory.
\end{abstract}


\section{Introduction}
\label{sec:intro}

Astrophysical fluids are often compressible, stratified, magnetized and multiphase, with cooling and heating processes spanning many orders of magnitude in temperature and density. These properties give rise to a broad spectrum of hydrodynamic and magnetohydrodynamic instabilities that govern many fundamental astrophysical processes. The Rayleigh-Taylor (RT) instability \citep{Rayleigh1883,Taylor1950} and Kelvin-Helmholtz (KH) instability \citep{Chandrasekhar1961} drive buoyancy-induced overturning and shear-induced mixing, while thermal instability \citep{field1965thermal} shape the multiphase structure of the interstellar and circumgalactic medium \citep{faucher2023key}. The magnetorotational instability (MRI, \citealt{balbus1991powerful}) drives angular momentum transport in accretion disks; the Parker instability \citep{parker1958dynamical} evacuates gas from galactic disks; and the baroclinic vorticity generation ($\propto\nabla p\times\nabla\rho$) initiates the Richtmyer-Meshkov instability \citep{richtmyer1960taylor} at shocked interfaces.

Classical linear stability theory describes the long-term asymptotic behavior of infinitesimal perturbations through the spectrum of the linearized operator, with the spectral radius determining whether perturbations grow or decay asymptotically \citep{Chandrasekhar1961}. In many applications, however, the onset and finite-time amplification of perturbations are of greater physical relevance than their asymptotic behavior. Although the evolution of individual eigenmodes is well understood, substantial transient growth can arise from the constructive interaction of multiple non-normal modes, even when all eigenmodes are asymptotically stable. Pseudospectral analysis provides a qualitative understanding of this non-modal growth mechanism \citep{ButlerFarrell1992,Trefethen1993,Schmid2007}.  However, it neither quantitatively characterizes the maximum finite-time amplification nor identifies the initial perturbation responsible for achieving it. 

The aforementioned limitations raise a fundamental question common to all instability problems: among all possible initial perturbations of equal energy, which spatial structure is the most dangerous -- that is, which seed triggers the greatest finite-time amplification in the subsequent nonlinear evolution? Answering this question motivates the development of optimal perturbation theory. To quantitatively identify perturbations responsible for the greatest finite-time amplification, \citet{Mu2003} introduced the conditional nonlinear optimal perturbation (CNOP) approach. By formulating the perturbation growth problem as a nonlinear optimization problem, CNOP numerically determines the small-amplitude initial perturbation that leads to the maximum finite-time amplification under a prescribed constraint. The CNOP approach has been successfully applied in geophysical fluid dynamics, including studies of El~Ni\~{n}o predictability and climate sensitivity \citep{MuDuan2003,Mu2004,Duan2009}. Subsequently, this methodology has been extended to broader fluid dynamics applications. In transitional shear flows, the same idea is commonly referred to as nonlinear optimal perturbation theory \citep{PringleKerswell2010,Monokrousos2011,Pringle2012,Kerswell2018}, and similar techniques have also been utilized in thermoacoustic system \citep{Juniper2011}. More recently, sampling-based methods have been developed to reduce the computational cost associated with adjoint calculations \citep{shi2023adjoint}. Furthermore, by exploiting parallel computation, these sampling approaches can be efficiently implemented for large-scale nonlinear optimal perturbation problems \citep{shi2024sampling}.

Previous works have largely focused on single-phase, incompressible fluids without magnetic fields. In contrast, astrophysical environments are inherently complex multiscale and multiphysics systems, where shocks, radiative cooling, phase transitions, and magnetic reconnection may occur simultaneously. Understanding how nonlinear effects and the spatial structures of initial perturbations influence instability development in such environments is therefore of significant importance, yet remains a major challenge. The RT instability provides an ideal test problem because it is common in astrophysical systems, possesses a well-established linear theory, and exhibits strong nonlinear sensitivity to initial conditions. In the linear regime, RT growth proceeds at the rate $\sigma = \sqrt{A\,g\,k}$ for Atwood number $A$, gravity $g$, and wavenumber $k$ \citep{Chandrasekhar1961}. However, the subsequent nonlinear turbulent regime is more complicated. The Alpha-Group collaboration \citep{Dimonte2004}, based on a suite of high-resolution three-dimensional simulations with different numerical methods and comparisons with laboratory experiments, showed that the RT growth  depends strongly on the initial mode spectrum: a seed with only short-wavelength modes gives a growth rate significantly below the experimental value. This result indicates nonlinear sensitivity to the spatial organization of the initial perturbation, which is beyond the scope of linear analysis and is exactly the type of sensitivity that CNOP is designed to quantify. In astrophysics, RT instability governs the deceleration of supernova ejecta into circumstellar shells \citep{Chevalier1976,Kifonidis2003}, convective boundary mixing in stellar interiors \citep{Meakin2007}, compositional mixing at the core-mantle boundary of proto-neutron stars \citep{Burrows1995}, and the inner ring structure of young supernova remnants \citep{Jun1995,Wang2001}. In all of these settings, the seed perturbation is set by processes that are difficult to constrain precisely.

In this paper, we present the first application of CNOP to an astrophysical fluid instability, using the two-dimensional compressible RT instability as our test case. We demonstrate that spatially structured optimal perturbations trigger RT-driven turbulent mixing much more efficiently than unstructured random perturbations with the same energy budget. In controlled comparisons, the CNOP perturbations produce a substantially larger kinetic-energy response than random perturbations. We discuss the broader implications of these findings for astrophysical observations and theoretical modeling in~\S\ref{sec:conclusions}.

The paper is organized as follows. In \S\ref{sec:methods}, we describe the governing equations, initial conditions and numerical setup. In \S\ref{sec:cnop}, we formulate the CNOP problem and describe the optimization algorithm. We present the optimal perturbations and their forward nonlinear evolution in \S\ref{sec:results}, and finally discuss and conclude in \S\ref{sec:conclusions}. Convergence diagnostics for the CNOP optimizer are collected in Appendix~\ref{app:convergence}, and the two-dimensional Fourier reconstruction is presented in Appendix~\ref{app:2d_fourier}.

\section{RT Instability Setup}
\label{sec:methods}

\subsection{Governing Equations}
\label{sec:rt_physics}

We model the fluid as a compressible ideal gas described by the two-dimensional Euler equations with a uniform external gravitational field. In conservative form,
\begin{align}
  \frac{\partial\rho}{\partial t}
  + \nabla\cdot(\rho\bm{u}) &= 0,
  \label{eq:mass}
  \\[4pt]
  \frac{\partial(\rho\bm{u})}{\partial t}
  + \nabla\cdot(\rho\bm{u}\bm{u} + p\bm{I})
  &= \rho\bm{g},
  \label{eq:momentum}
  \\[4pt]
  \frac{\partial E}{\partial t}
  + \nabla\cdot\bigl[(E + p)\bm{u}\bigr]
  &= \rho\bm{u}\cdot\bm{g},
  \label{eq:energy}
\end{align}
where $\rho$ is the mass density, $\bm{u} = (u_x, u_y)$ is the velocity field, $p$ is the thermal pressure, $\bm{I}$ is the identity tensor, $\bm{g} = -g\hat{y}$ is the gravitational acceleration (pointing in the $-y$ direction), and
\begin{equation}
  E = \frac{p}{\gamma - 1} + \frac{1}{2}\rho\left|\bm{u}\right|^2
  \label{eq:total_energy}
\end{equation}
is the total energy density per unit volume. The ideal-gas equation of state $p = (\gamma - 1)\rho e$ closes the system, where $e$ is the specific internal energy. Throughout this work we take $\gamma = 5/3$, appropriate for a monatomic ideal gas.

The domain is the unit square $[0,L]\times[0,L]$ with $L=1$. Boundary conditions are periodic in $x$ and hydrostatic in $y$ ($\partial p/\partial y = -\rho g$).

\subsection{Initial Conditions and Linear Stability}
\label{sec:ics}

The initial unperturbed configuration is a two-layer stratified atmosphere in hydrostatic equilibrium. The density profile is smoothed across the interface to avoid Gibbs phenomena on the numerical grid,
\begin{equation}
  \rho_0(y) = \frac{\rho_h + \rho_l}{2}
  + \frac{\rho_h - \rho_l}{2}
    \tanh\!\left(\frac{y - y_0}{h_s}\right),
  \label{eq:rho0}
\end{equation}
where $y_0 = L/2$ is the interface position, $h_s = 0.02\,L$ is the smoothing length, and $\rho_h = 2$, $\rho_l = 1$ are the heavy (top) and light (bottom) densities. The Atwood number is $A = (\rho_h - \rho_l)/(\rho_h + \rho_l) = 1/3$. The pressure profile is obtained by integrating the hydrostatic balance equation $\partial p/\partial y = -\rho g$ with the boundary condition $p(y=L) = \rho_h g L$. The initial velocity field is zero except for the perturbation $u_{y,0}(\bm{r})$ described in \S\ref{sec:cnop_problem}.

For the smoothed profile, the classical RT growth rate $\sigma_k = \sqrt{A\,g\,k}$ is modified by the finite transition layer thickness. In the thin-interface limit $k h_s \ll 1$, the growth rate reduces to
\begin{align}
  \sigma_k = \sqrt{A\, g\, k}, \quad
  \tau_{\rm RT} = \sigma_k^{-1}
  = \left(A\, g\, k\right)^{-1/2},
  \label{eq:sigma}
\end{align}
where $k = 2\pi n_k / L$ for mode number $n_k$. For the reference parameters used below ($n_k = 2$, $g = 1$, $L = 1$), the linear e-folding time is $\tau_{\rm RT} = (A\, g\, k)^{-1/2} \sim 0.49$.

\subsection{Numerical Implementation}
\label{sec:implementation}

The governing equations are solved on a uniform $N \times N$ Cartesian grid using a finite-volume Godunov-type scheme with a hybrid HLLC Riemann solver \citep{Toro1994} and a gravitational source term. The unperturbed baseline run provides the reference response that is subtracted in the objective function in \S\ref{sec:cnop_problem}.

The solver is implemented with a customized version of Astronomix \citep{storcks_astronomix_2025}, a Python library for astrophysical hydrodynamics built on JAX \citep{Bradbury2018}. The JAX framework makes the entire discrete time-marching sequence differentiable through source-code transformation. This allows us to compute exact gradients of the discretized objective with respect to the initial perturbation using reverse-mode automatic differentiation, without hand-coded adjoint equations.

\section{CNOP Formulation}
\label{sec:cnop}

\subsection{Problem Statement}
\label{sec:cnop_problem}

Let $F_t$ denote the nonlinear forward map that advances the full compressible state from time $0$ to time $t$ starting from the equilibrium configuration. To simplify notation, fields evaluated at the initial time $t=0$ are denoted by a subscript~0 (e.g. $\rho_0$, $u_{x,0}$, $e_{\mathrm{k},0}$), while fields evaluated at the optimization horizon $t^*$ are denoted by a superscript~* (e.g. $e_{\mathrm{k}}^*\equiv e_{\mathrm{k}}(t^*)$). We restrict perturbations to the $y$-component of velocity, so the perturbation field is $\delta u_0(\bm r)\equiv u_{y,0}(\bm r)$. Here $\bm{r}=(x,y)$ denotes the two-dimensional position vector. All other fields ($\rho$, $u_x$, $p$) retain their hydrostatic values at $t=0$.

Denoting the kinetic-energy density of the final state as
\begin{equation}
  e_{\mathrm{k}}(\bm{r}, t) = \frac{1}{2}\rho(\bm{r},t)
                \left[u_x^2(\bm{r},t) + u_y^2(\bm{r},t)\right],
  \label{eq:ke}
\end{equation}
we define the scalar objective function
\begin{equation}
  J(\delta u_0) = \int_{\Omega}
  \left[e_{\mathrm{k}}^*(\bm{r};\, \delta u_0) - e_{\mathrm{k}}^*(\bm{r};\, 0)\right]^2
  \,dV,
  \label{eq:objective}
\end{equation}
where $\Omega$ is the two-dimensional domain and $dV = dx\,dy$. This objective measures the squared departure of the kinetic-energy field from the baseline. Maximizing $J$ therefore searches for perturbations that produce the largest kinetic-energy anomaly at the target time $t^*$.

The initial perturbation is constrained by a kinetic-energy budget expressed in terms of the initial perturbation energy,
\begin{equation}
  C(\delta u_0) \equiv
  \int_{\Omega} \frac{1}{2}\rho_0(\bm{r})\, \delta u_0^2(\bm{r})\,dV
  \leq \varepsilon.
  \label{eq:constraint}
\end{equation}
This kinetic-energy constraint enforces a fixed energy budget for the seed perturbation, independent of the grid resolution $N$.

The CNOP problem is to find the optimal initial velocity perturbation $\delta u_{0,\mathrm{opt}}$ that maximizes $J$ subject to the constraint $C(\delta u_0) \leq \varepsilon$:
\begin{equation}
\label{eqn: cnop}
  \delta u_{0,\mathrm{opt}}  = \mathop{\arg\max}_{C(\delta u_0)\leq\varepsilon}\, J(\delta u_0).
\end{equation}
Since the optimizers are implemented as minimizers, we solve the equivalent minimization problem for $-J(\delta u_0)$.
\subsection{Interpretation}
\label{sec:interpretation}

Eq.~\eqref{eq:objective} measures the growth beyond what the unperturbed system would produce. For the RT problem, the unperturbed configuration remains close to static, so the baseline kinetic-energy density is small. In practice, the subtraction in Eq.~\eqref{eq:objective} removes the residual numerical response of the smoothed atmosphere, and $J(\delta u_0)$ is therefore dominated by the kinetic-energy field generated by the perturbation itself: $J(\delta u_0) \sim \int_\Omega \left[e_{\mathrm{k}}^*(\bm{r}; \delta u_0)\right]^2\,dV$. Thus, the CNOP searches for the initial velocity perturbation that leads to the most energetic response in the flow at the target time $t^*$.

The choice of target time $t^*$ determines the regime of the problem. For $t^* \ll \tau_{\rm RT}$, the dynamics are approximately linear and the CNOP solution should be close to the fastest-growing normal mode. For $t^* \gg \tau_{\rm RT}$, the optimal perturbation can make use of nonlinear mechanisms that are absent from linearized analysis.

\subsection{Objective Function Choice}
\label{sec:cnop_objective}

We use the $L^2$ norm of the $e_{\mathrm{k}}$ field difference as the objective (Eq.~\ref{eq:objective}), rather than a norm of the state difference as in the original CNOP formulation \citep{Mu2003}. $e_{\mathrm{k}}$ directly quantifies the conversion of gravitational potential energy into mechanical motion, which is the central diagnostic for RT growth. The field-level $L^2$ norm includes both the total amplitude and the spatial coherence of the kinetic-energy anomaly, and therefore favors perturbations that produce structured, large-scale RT overturning rather than only local velocity fluctuations.

\subsection{Gradient Computation and Optimization}
\label{sec:grad_opt}

Computing the gradient $\nabla J$ with respect to the perturbation field by finite differences would require $O(N^2)$ independent forward simulations, which is prohibitively expensive for multi-dimensional problems. Since the entire Astronomix time-marching sequence is implemented in JAX, a single reverse-mode automatic differentiation pass gives the exact gradient at a cost comparable to $3$--$5\times$ a single forward run, independent of $N$ \citep{Bradbury2018,storcks_astronomix_2025}. Memory is managed with binomial checkpointing \citep{Griewank2000}, which avoids storing all $N_t \sim 10^3$ intermediate states by re-computing short segments from a sparse set of saved checkpoints. This keeps the peak memory overhead manageable, with negligible runtime cost for the problem sizes considered here.

The CNOP is found with a projected L-BFGS quasi-Newton algorithm \citep{NocedalWright2006}. At each iteration, a descent direction is constructed from the 10 most recent displacement-gradient pairs, which implicitly approximate the inverse Hessian. The trial step is then rescaled to satisfy $C(\delta u_0)\leq\varepsilon$ by
\begin{equation}
  P(\delta u_0) =
  \begin{cases}
    \delta u_0, & C(\delta u_0) \leq \varepsilon,\\[4pt]
    \delta u_0\,\displaystyle\left(\frac{\varepsilon}{C(\delta u_0)}\right)^{1/2}, & \text{otherwise},
  \end{cases}
  \label{eq:projection}
\end{equation}
where $P$ denotes the energy-constraint projection. This mapping exactly saturates the energy budget when the constraint is active. The step length is determined by a non-monotone Armijo backtracking line search \citep{GriLoLaUsTho1986}, which permits occasional temporary increases in objective value to help avoid shallow local traps. Convergence diagnostics are presented in Appendix~\ref{app:convergence}.

The initial guess is a zero-mean Gaussian random velocity field localized near the density interface (width $h_{\rm loc} = 0.1\,L$), projected to satisfy $C(\delta u_0) = \varepsilon$. This choice is deliberately unstructured and sub-optimal: as the forward comparison runs in \S\ref{sec:comp_runs} show, such a random perturbation produces only weak RT growth. The optimizer nevertheless converges from this starting point to a much more effective CNOP solution. We have verified that the CNOP solution is insensitive to the initial random seed: different seeds converge to approximately the same objective value and qualitatively identical spatial structure.

\subsection{Parameter Choices}
\label{sec:params}

Tab.~\ref{tab:params} lists the physical and numerical parameters shared by all runs. We explore three combinations of grid resolution $N$ and optimization horizon $t^*$, summarized in Tab.~\ref{tab:runs}; each combination is assigned a shorthand label used throughout the text.

\begin{table}[h]
\centering
\caption{Fixed parameters used across all simulations.}
\label{tab:params}
\begin{tabular}{lll}
\hline\hline
\multicolumn{3}{c}{RT-related parameters}\\
\hline
Domain size & $L$ & $1$ \\
Heavy density & $\rho_h$ & $2$ \\
Light density & $\rho_l$ & $1$ \\
Atwood number & $A$ & $1/3$ \\
Gravitational acceleration & $g$ & $1$ \\
\hline
\multicolumn{3}{c}{CNOP-related parameters}\\
\hline
Perturbation constraint & $\varepsilon$ & $10^{-6}$ \\
L-BFGS history length & $m$ & $10$ \\
\hline
\end{tabular}
\end{table}

\begin{table}[t]
\centering
\caption{The three converged CNOP runs. Here $\tau_{\rm RT} = (A\,g\,k)^{-1/2} \sim 0.49$ is the linear RT e-folding time for the reference mode $n_k=2$.}
\label{tab:runs}
\begin{tabular}{lccccc}
\hline\hline
Label & $N$ & $t^*$ & $t^*/\tau_{\rm RT}$ & Max iteration \\
\hline
\texttt{T2\_N128} & $128$ & $2.0$ & $4.1$ & 200 \\
\texttt{T2\_N256} & $256$ & $2.0$ & $4.1$ & 3000 \\
\texttt{T3\_N128} & $128$ & $3.0$ & $6.1$ & 5800 \\
\hline
\end{tabular}
\end{table}

The constraint $\varepsilon = 10^{-6}$ is calibrated so that, for a uniform $u_y$ perturbation, it corresponds to an rms velocity $\langle u_y^2\rangle^{1/2} \sim (\varepsilon / \langle\rho_0\rangle N^2\Delta x\,\Delta y)^{1/2} \sim 3\times10^{-4}$, which is well within the linear regime of RT instability at the initial time ($t=0$).

\subsection{Forward Comparison Run Design}
\label{sec:comp_runs}

To compare each CNOP solution with physically motivated alternatives, we associate every converged CNOP run ({\tt T2\_N128}, {\tt T2\_N256}, and {\tt T3\_N128}) with a suite of forward RT simulations. The energy budget of the initial perturbation is fixed at $\varepsilon=10^{-6}$ throughout. Runs within each group differ \emph{only} in the choice of initial velocity perturbation $u_{y,0}(\bm{r})$. This design provides two controlled comparisons: (i) {\tt T2\_N128} versus {\tt T2\_N256} isolates the effect of spatial resolution at fixed target time $t^*=2.0$, and (ii) {\tt T2\_N128} versus {\tt T3\_N128} isolates the effect of the optimization horizon at fixed resolution $N=128$. We consider three perturbation types:

\begin{itemize}[nosep]
  \item \textbf{\pCNOP} --- the optimal perturbation $\delta u_{0,\mathrm{opt}}$ obtained directly from the corresponding CNOP optimization run ({\tt T2\_N128}, {\tt T2\_N256}, or {\tt T3\_N128}).
  \item \textbf{\pRand} --- a zero-mean Gaussian random field localized near the density interface (width $h_{\rm loc}=0.1\,L$), normalized so that $C(\delta u_0)=\varepsilon$. This is identical in construction to the initial guess used by the CNOP optimizer (\S\ref{sec:grad_opt}), so the comparison between \pRand\ and \pCNOP\ directly quantifies how much the optimization improves the initial perturbation.
  \item \textbf{\pMmode} --- a reduced perturbation reconstructed from the five largest Fourier modes of the corresponding CNOP seed, retaining their relative amplitudes and phases. Its construction is described in \S\ref{sec:fourier_structure}.
\end{itemize}

All nine forward comparison runs are summarized in Tab.~\ref{tab:comp_runs}. The structure of the \pCNOP\ perturbation and its objective value $J(\delta u_{0,\mathrm{opt}})$ are determined by the corresponding CNOP optimization (\S\ref{sec:cnop}). The forward simulations then re-evolve the system from $t=0$ to the corresponding target-time interval with uniformly spaced diagnostic snapshots. This allows us to examine how the gain optimized at $t^*$ appears in the nonlinear RT evolution.

\begin{table*}
\centering
\caption{%
  Forward RT comparison runs. The energy constraint $\varepsilon=10^{-6}$ is fixed for all runs. The ``Source CNOP run'' column identifies the corresponding CNOP optimization (Tab.~\ref{tab:runs}); \pMmode\ is constructed from the five dominant Fourier components of the corresponding CNOP, whereas \pRand\ is generated independently. The upper group ({\tt T2\_N128} and {\tt T2\_N256}) compares the effect of spatial resolution at fixed $t^*=2.0$; the lower group ({\tt T3\_N128}) is compared with {\tt T2\_N128} isolate the effect of the target time (i.e., the optimization time horizon).}
\label{tab:comp_runs}
\begin{tabular}{llccl}
\hline\hline
Run label & Source CNOP run & $N$ & $t^*$ & Initial perturbation $u_{y,0}$ \\
\hline
\multicolumn{5}{l}{\textit{Resolution comparison at $t^*=2.0$: {\tt T2\_N128} vs.\ {\tt T2\_N256}}} \\
\texttt{T2\_N128-CNOP}    & {\tt T2\_N128} & 128 & 2.0 & \parbox[t]{5.5cm}{Optimal $\delta u_{0,\mathrm{opt}}$ from {\tt T2\_N128}} \\
\texttt{T2\_N128-Random}  & {\tt T2\_N128} & 128 & 2.0 & \parbox[t]{5.5cm}{Random Gaussian, normalized to $C=\varepsilon$} \\
\texttt{T2\_N128-5modes}  & {\tt T2\_N128} & 128 & 2.0 & \parbox[t]{5.5cm}{Five dominant CNOP Fourier modes} \\
\texttt{T2\_N256-CNOP}    & {\tt T2\_N256} & 256 & 2.0 & \parbox[t]{5.5cm}{Optimal $\delta u_{0,\mathrm{opt}}$ from {\tt T2\_N256}} \\
\texttt{T2\_N256-Random}  & {\tt T2\_N256} & 256 & 2.0 & \parbox[t]{5.5cm}{Random Gaussian, normalized to $C=\varepsilon$} \\
\texttt{T2\_N256-5modes}  & {\tt T2\_N256} & 256 & 2.0 & \parbox[t]{5.5cm}{Five dominant CNOP Fourier modes} \\
\hline
\multicolumn{5}{l}{\textit{Target-time comparison at $N=128$: {\tt T2\_N128} vs.\ {\tt T3\_N128}}} \\
\texttt{T3\_N128-CNOP}    & {\tt T3\_N128} & 128 & 3.0 & \parbox[t]{5.5cm}{Optimal $\delta u_{0,\mathrm{opt}}$ from {\tt T3\_N128}} \\
\texttt{T3\_N128-Random}  & {\tt T3\_N128} & 128 & 3.0 & \parbox[t]{5.5cm}{Random Gaussian, normalized to $C=\varepsilon$} \\
\texttt{T3\_N128-5modes}  & {\tt T3\_N128} & 128 & 3.0 & \parbox[t]{5.5cm}{Five dominant CNOP Fourier modes} \\
\hline
\end{tabular}
\end{table*}

\section{Results}
\label{sec:results}

\subsection{The CNOP Seed Is Coherent and Interface-Localized}
\label{sec:cnop_structure}

\begin{figure*}[!ht]
  \centering
  \includegraphics[width=\textwidth]{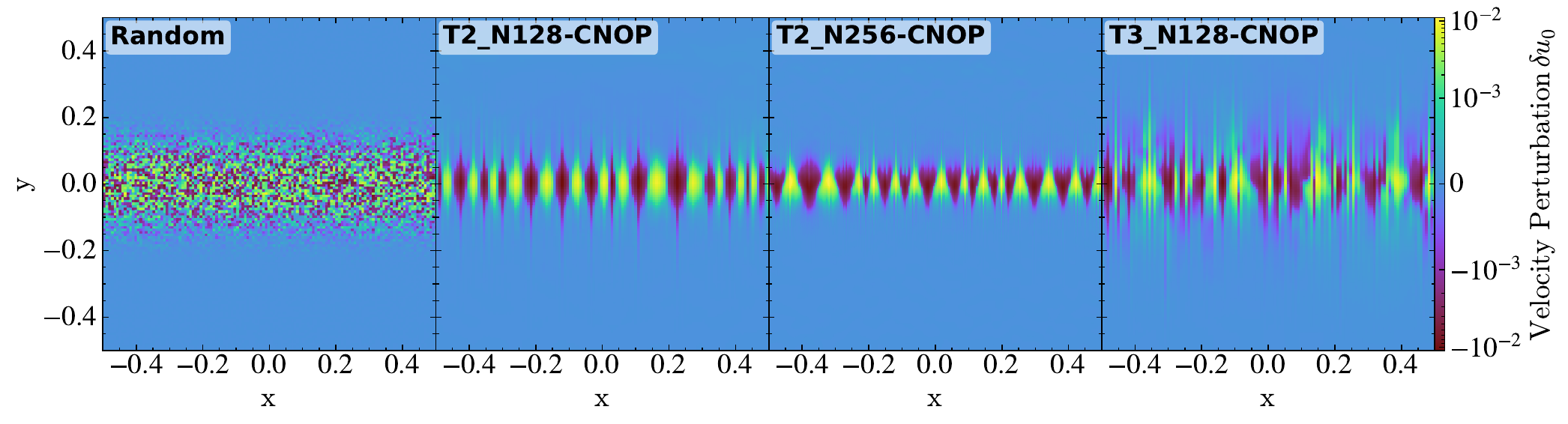}
  \caption{Initial vertical-velocity perturbations. The left panel shows the localized random field used as the initial guess. The remaining panels show the CNOPs obtained by solving the optimization problem\eqref{eqn: cnop} for the cases: {\tt T2\_N128}, {\tt T2\_N256}, and {\tt T3\_N128}. All perturbations satisfy the same kinetic-energy constraint, $C(\delta u_0)=10^{-6}$.}
  \label{fig:cnop_structure}
\end{figure*}

Fig.~\ref{fig:cnop_structure} shows the CNOPs obtained by solving the optimization problem stated in Eq.~\eqref{eqn: cnop}, together with an incoherent localized random field used as the initial guess for comparison. In contrast to the incoherent random field, the CNOPs exhibit an coherent wave packet structure concentrated around the unstable interface. Comparing the two $t^*=2$ solutions, the wave packet becomes slightly more localized as the spatial resolution increases, although the difference is modest. The close argeement between {\tt T2\_N128} and {\tt T2\_N256} indicates that the CNOP has already reached good convergence with respect to spatial resolution and that the observed structure is not a numerical artifact.

The longer-horizon solution, {\tt T3\_N128}, is visibly different. It features broader vertical streaks and a less periodic horizontal structure, while still concentrating most of its perturbation energy near the unstable interface. This behavior is expected because a longer optimization horizon allows the initial perturbation to exploit nonlinear interactions and phase reorganization, rather than relying solely on the initial exponential growth. Consequently, the CNOP depends on the target time $t^*$, as expected for a nonlinear predictability problem.

\subsection{Spectral Structure of the CNOP Seeds}
\label{sec:fourier_structure}

\begin{figure*}[!ht]
  \centering
  \includegraphics[width=\textwidth]{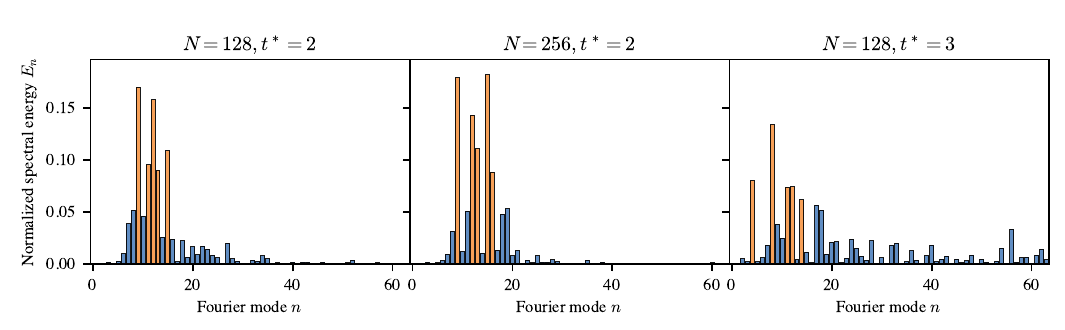}
  \caption{Normalized spectral energy $E_n$ of CNOPs across the horizontal Fourier modes $n$. The orange bars indicate the five dominant modes used to construct the \pMmode\ comparison runs, whereas the blue bars represent the remaining Fourier modes.}
  \label{fig:fourier}
\end{figure*}

To further characterize the CNOP seeds,  we analyze their Fourier spectra obtained via the fast Fourier transform (FFT).  For clarity, we focus on the one-dimensional horizontal Fourier decomposition, which provides a clearer view of the spectral distribution, while the more comprehensive two-dimensional Fourier analysis is presented in Appendix~\ref{app:2d_fourier}. Specifically, we first identify the row $y_{\rm peak}$, where the root-mean-square (rms) value of $u_{y,0}$ along the $x$-direction attains its maximum. The corresponding horizontal Fourier coefficients $\hat{u}_n$, together with their normalized spectral energy $E_n$, are then computed as:
\begin{subequations}
\label{eq:fourier_coeff}
\begin{empheq}[left=\empheqlbrace]{align}
        \hat{u}_n &= \sum_{j=0}^{N-1}u_{y,0}(x_j,y_{\rm peak})\, \mathrm{exp}\left(\frac{-2\pi i n j}{N}\right),    \label{eq:fourier_coeff_u} \\
       E_n &= \frac{|\hat{u}_n|^2}{\sum_m |\hat{u}_m|^2},
       \label{eq:fourier_coeff_energy} 
\end{empheq}     
\end{subequations}
where $E_n$ is proportional to the kinetic energy of the velocity perturbation carried by mode $n$. Since the perturbation wave is real-valued, the Fourier spectrum is symmetric with respect to positive and negative wavenumbers. Therefore, only the positive wavenumbers are presented, and the normalization in Eq.~\eqref{eq:fourier_coeff_energy} is applied over the plotted positive modes.  The normalized spectral energies $E_n$ for the CNOPs obtained for the cases, {\tt T2\_N128}, {\tt T2\_N256}, and {\tt T3\_N128}, are shown in Fig.~\ref{fig:fourier}, illustrating how the perturbation energy is distributed among the horizontal Fourier modes.
 
Fig.~\ref{fig:fourier} provides a concise description of the spectral-energy distribution. For the two cases with the short time horizon ($t^*=2$), specifically {\tt T2\_N128}, {\tt T2\_N256}, the dominant energy is concentrated within a relatively narrow group: $n \sim 9$--$16$. As the grid resolution increases from $N=128$ to $N=256$, this concentration becomes more pronounced.  In other words, when the time horizon is short, the perturbation energy is highly localized in Fourier space, which demonstrates that the CNOPs obtained over short time horizons closely approximate the linear regime. In contrast, comparing the two cases with the same resolution $N=128$, specifically {\tt T2\_N128} and {\tt T3\_N128}, the spectrum is more dispersed for the longer time horizon ($t^*=3$) than that for the shorter one ($t^*=2$). This suggests that as more modes participate in the construction of the CNOPs for the longer time horizons, a greater number of non-normal interactions are engaged, reflecting the increasing influence of nonlinear effects.  
To further support these findings, a two-dimensional Fourier reconstruction, retaining 90\% of the total Fourier spectral energy, is provided in Appendix~\ref{app:2d_fourier}.

\subsection{Total Kinetic-Energy Growth}
\label{sec:ek_growth_section}

\begin{figure*}
  \centering
  \includegraphics[width=\textwidth]{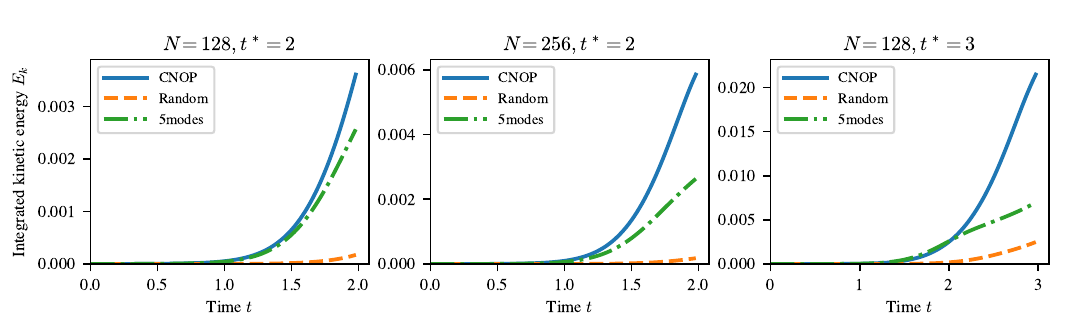}
  \caption{The comparison of total kinetic energy growth generated by CNOPs.}
  \label{fig:ek_growth}
\end{figure*}

In the following two subsections, we investigate the nonlinear growth induced by the CNOP seeds. Fig.~\ref{fig:ek_growth} compares the evolution of the total kinetic energy for the full CNOPs, the localized random perturbations shown in Fig.~\ref{fig:cnop_structure}, and the dominant \pMmode\ approximation of the CNOPs identified in Fig.~\ref{fig:fourier}. For the baseline case, {\tt T2\_N128}, the nonlinear evolution initiated by the dominant \pMmode\ approximation closely follows that of the full CNOP, whereas the localized random perturbation produces a markedly different evolution. As the grid resolution increases from $N=128$ to $N=256$, the discrepancy between the dominant \pMmode\ approximation and the full CNOP becomes more pronounced, indicating that nonlinear interactions involving the non-dominant Fourier modes become increasingly important at higher resolutions.

Keeping the grid resolution fixed while increasing the optimization time horizon from $t=2$ to $t=3$, we find that the dominant \pMmode\ approximation remains in close agreement with the full CNOP during the early stage ($t<2$). However, over the later interval ($2\le t\le3$), the discrepancy grows rapidly. This behavior indicates that nonlinear interactions between the dominant and non-dominant Fourier modes become increasingly significant over longer time horizons. As the evolution departs further from the near-linear regime, the dynamics can no longer be accurately represented by only the dominant Fourier modes. This phenomenon is even more evident in the two-dimensional Fourier reconstruction retaining 90\% of the total Fourier spectral energy, as shown in Appendix~\ref{app:2d_fourier}.

\subsection{Mixing-layer Growth}
\label{sec:mixing_width}

While the kinetic energy characterizes the overall mechanical response of the flow, astrophysical applications are often more concerned with the efficiency and extent of turbulent mixing. To quantify this process, we evaluate the growth of the mixing layer using the standard diagnostic based on the horizontally averaged density:
\begin{equation}
  h_{\rm mix}(t) =
  \int_0^L 4 f(y,t) \left[1-f(y,t)\right] dy,\\
  \label{eq:hmix}
\end{equation}
where $f(y,t)$ is the normalized density profile, defined as:
\begin{equation}
  \label{eq:hmix-f}
 f(y,t) = \mathrm{clip}\!\left[
  \frac{\langle\rho\rangle_x-\rho_l}{\rho_h-\rho_l},0,1\right]
\end{equation}
and $\langle \rho \rangle_x$ denotes the horizontal average of the density field. The clipping operator restricts the normalized density to the interval $[0, 1]$.  This diagnostic is widely used in astrophysical fluid dynamics to quantify the broadening of the interface between the two fluids. The integrand $4f(1-f)$ is zero in the unmixed pure-fluid regions and attains its maximum value at a fully mixed interface, so $h_{\text{mix}}(t)$  provides a robust measure of the width of the mixing layer. We use this quantity to compare the mixing efficiency of the full CNOPs, the localized random perturbations, and the dominant \pMmode\ approximation.

\begin{figure*}[!ht]
  \centering
  \includegraphics[width=\textwidth]{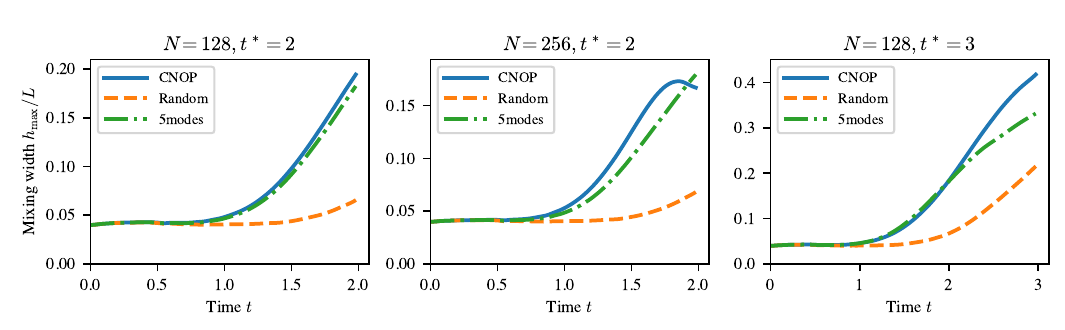}
  \caption{The comparison of mixing-layer growth generated by CNOPs with the mixing width given by Eq.~\eqref{eq:hmix}.}
  \label{fig:mixing_width}
\end{figure*}

Consistent with Fig.~\ref{fig:ek_growth}, Fig.~\ref{fig:mixing_width}  compares the nonlinear evolution of the mixing-layer thickness for the full CNOPs, the localized random perturbations shown in Fig.~\ref{fig:cnop_structure}, and the dominant \pMmode\ approximation of the CNOPs identified in Fig.~\ref{fig:fourier}. For the baseline case,  {\tt T2\_N128}, the mixing-layer evolution produced by the dominant \pMmode\ approximation is nearly indistinguishable from that of the full CNOP.  As the grid resolution increases from $N=128$ to $N=256$, a modest discrepancy emerges between the dominant \pMmode\ approximation and the full CNOP, although the two remain in close agreement throughout the evolution. This discrepancy is substantially smaller than that observed for the total kinetic energy, indicating that grid resolution is not the primary factor governing the evolution of the mixing layer. This observation is consistent with the spectral analysis in \S~\ref{sec:fourier_structure}, which showed that the dominant \pMmode\ approximation captures nearly the same fraction of the perturbation energy at both resolutions. In contrast, when the grid resolution is fixed and the optimization time horizon is extended from $t^*=2$ to $t^*=3$, the mixing-layer evolution predicted by the dominant \pMmode\ approximation progressively departs from that of the full CNOP. This trend indicates that nonlinear effects become increasingly important over longer optimization horizons. As the flow evolves further away from the near-linear regime, nonlinear interactions involving progressively more Fourier modes become significant, and the dynamics can no longer be adequately represented by only the dominant modes.

\subsection{Density Morphology}
\label{sec:morphology}

\begin{figure*}[!ht]
  \centering
  \begin{subfigure}[t]{\textwidth}
    \centering
    \includegraphics[width=\textwidth]{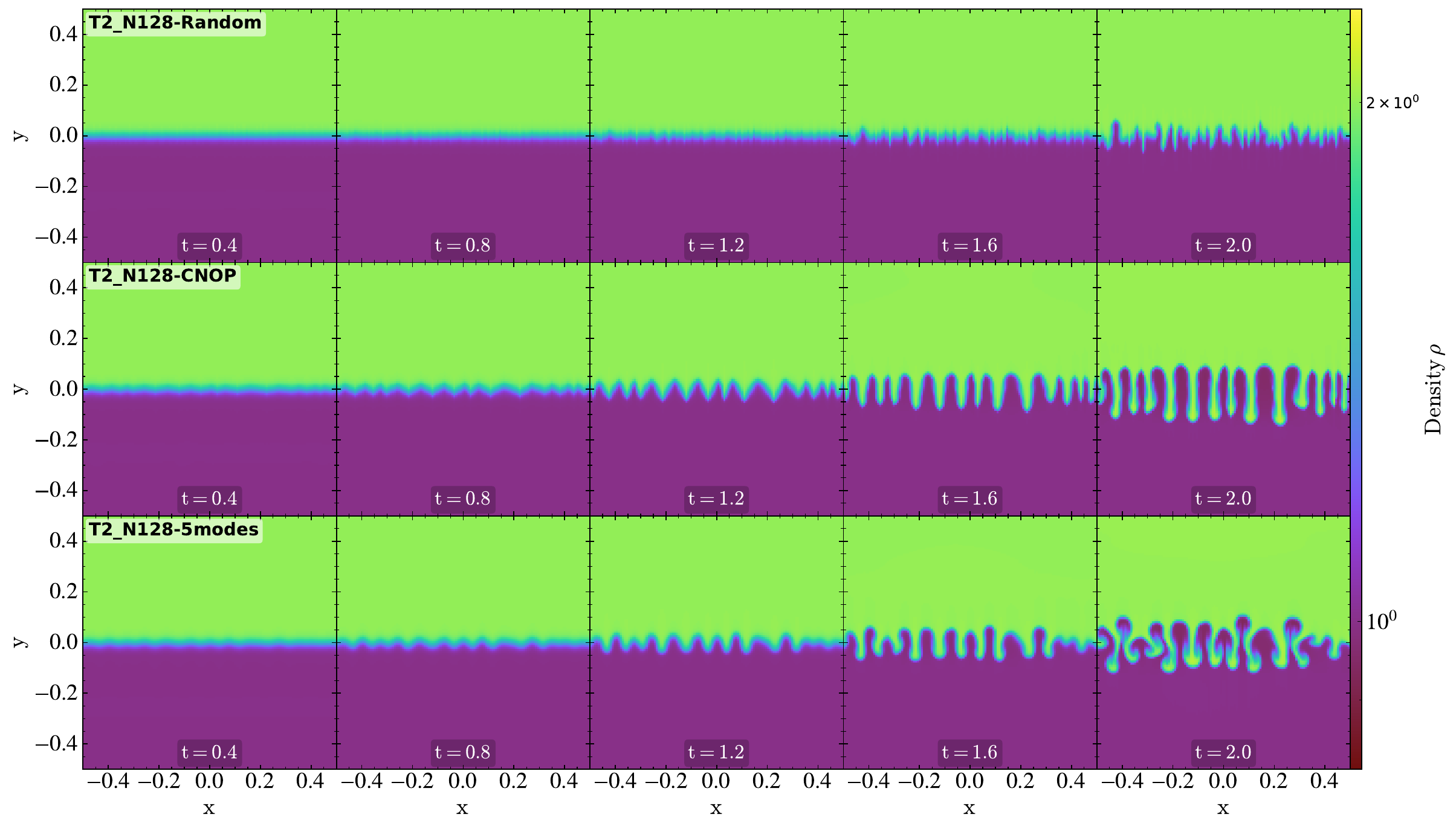}
    \caption{Density evolution for {\tt T2\_N128}.  The rows (from top to bottom) show the evolution initiated from the localized random perturbation, CNOP, and the dominant \pMmode\ reconstruction, respectively.}
    \label{fig:sim_evo_t2_n128}
  \end{subfigure}
  \vspace*{2cm}
  \begin{subfigure}[t]{\textwidth}
    \centering
    \includegraphics[width=\textwidth]{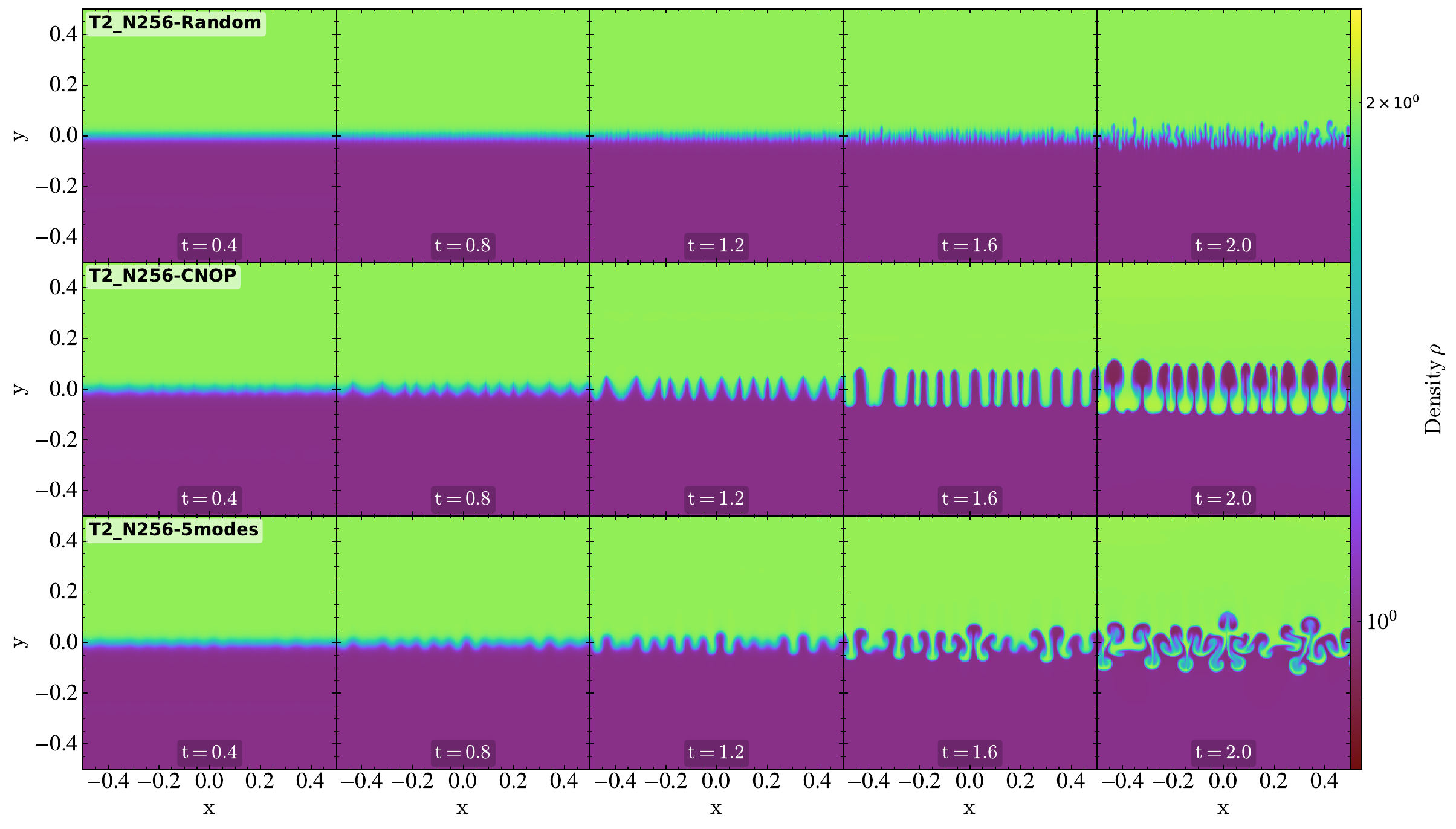}
    \caption{Density evolution for {\tt T2\_N256}, following the same row order as in Fig.~\ref{fig:sim_evo_t2_n128}.}
    \label{fig:sim_evo_t2_n256}
  \end{subfigure}
\end{figure*}
\begin{figure*}[!t]
  \ContinuedFloat
  \centering
  \begin{subfigure}[t]{\textwidth}
    \centering
    \includegraphics[width=\textwidth]{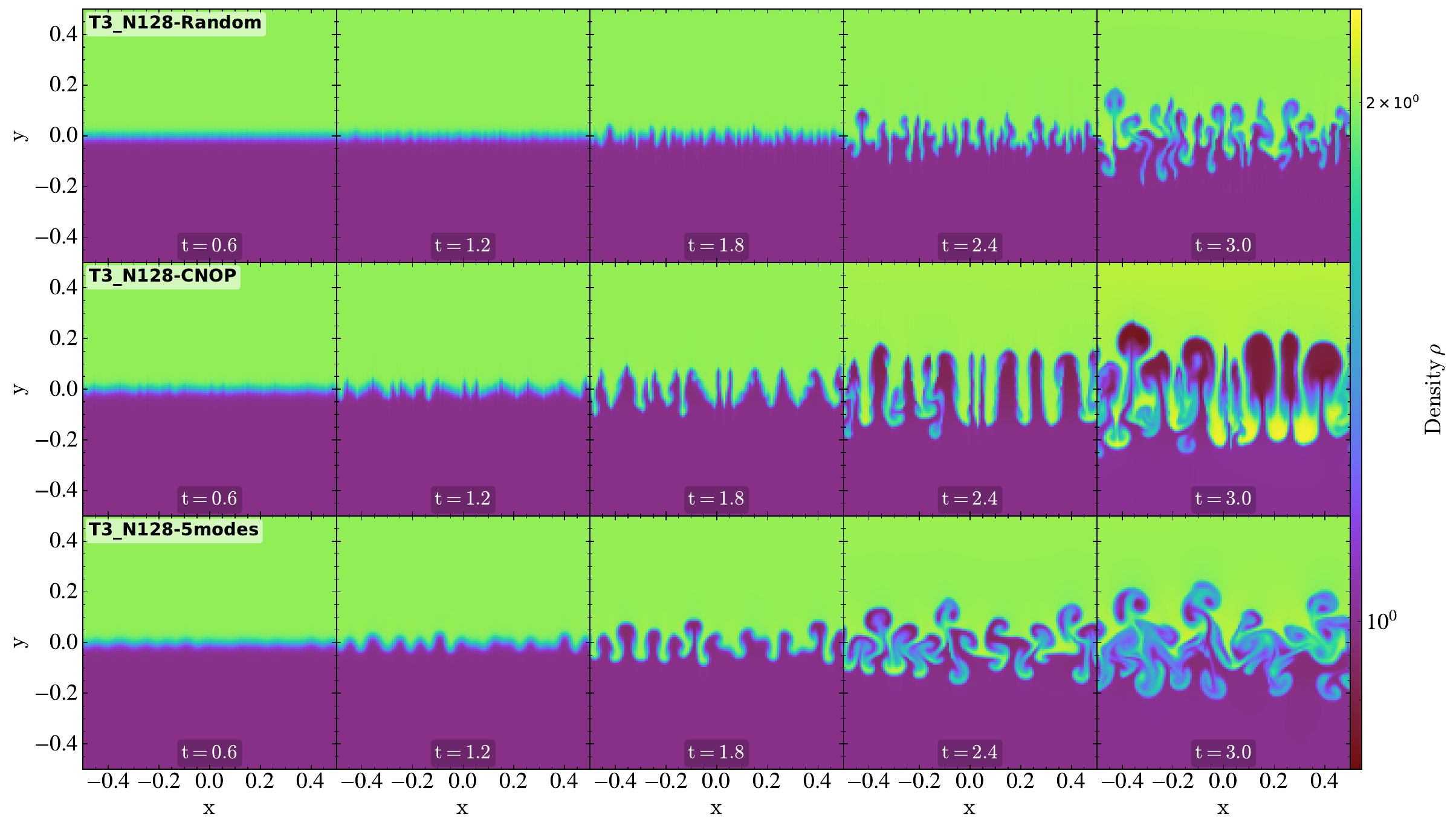}
    \caption{Density evolution for {\tt T3\_N128}, following the same row order as in Fig.~\ref{fig:sim_evo_t2_n128} and Fig.~\ref{fig:sim_evo_t2_n256}.}
    \label{fig:sim_evo_t3}
  \end{subfigure}
    \caption{Nonlinear evolution of the density field for the three initial perturbations: {\tt T2\_N128}, {\tt T2\_N256}, and {\tt T3\_N128}.}
    \label{fig:density_evo}
\end{figure*}

In this section, we synthesize the preceding analyses through the nonlinear evolution of the density field. This provides a unified view of how the different initial perturbations influence the subsequent development of the mixing layer, as illustrated in Fig.~\ref{fig:density_evo}.  For the baseline case, {\tt T2\_N128} (Fig.~\ref{fig:sim_evo_t2_n128}), the CNOP rapidly develops coherent nonlinear structures. By $t=1.2$, a train of bubbles and spikes with substantial interpenetration has already formed along the interface. As the evolution proceeds to $t=1.6$ and $t=2.0$, these structures continue to grow, merge, and intensify, resulting in increasingly vigorous overturning and mixing. In contrast, the localized random perturbation remains much closer to the initial interface throughout the same time interval. Although small-scale corrugations emerge, they lack the phase coherence required to trigger rapid, large-scale overturning, resulting in substantially weaker nonlinear development. The \pMmode\ approximation reproduces the coherent bubble-and-spike pattern at the early stage and subsequently develops nearly the same nonlinear roll-up and vortex-merging processes as the full CNOP. Consequently, it captures the principal density morphology throughout the entire evolution up to $t=2.0$, with only minor quantitative differences visible at the latest stage.

Fig.~\ref{fig:sim_evo_t2_n256} presents the higher-resolution case, {\tt T2\_N256}, allowing a comparison with the baseline resolution ($N=128$). As the grid resolution increases,  the CNOP develops a train of bubbles and spikes with pronounced interpenetration along the interface, and these coherent nonlinear structures become even more clearly resolved. The \pMmode\ approximation likewise reproduces the coherent bubble-and-spike pattern at the early stage and subsequently undergoes nearly the same nonlinear roll-up and vortex-merging processes as the full CNOP. Although the differences become slightly larger than in the lower-resolution case, the overall density morphology and temporal nonlinear evolution remain in close agreement, consistent with the kinetic-energy and mixing-width evolutions in Fig.~\ref{fig:ek_growth} and Fig.~\ref{fig:mixing_width}.

Fig.~\ref{fig:sim_evo_t3} present the longer-time case, {\tt T3\_N128}, at the same resolution.  Around the time $t=1.2$, a train of bubbles and spikes with substantial interpenetration has already formed along the interface.  However, as the flow evolves to $t=3.0$, the CNOP seed develops large rising bubbles, descending spikes, and secondary roll-up, leading to a broad and highly irregular mixing region with chaotic structures. The \pMmode\ approximation also enters a strongly nonlinear regime but generates a narrower mixing layer and does not reproduce the largest bubbles and spikes observed in the full CNOP. The random perturbation likewise becomes nonlinear, although the overturning occurs later and remains considerably weaker. These results support the spectral interpretation that a small set of dominant modes is sufficient to trigger organized nonlinear growth, whereas the remaining distributed modes of the t=3 optimum play an important role in the subsequent merging and late-time mixing. This behavior is also consistent with the kinetic-energy and mixing-width evolutions shown in Fig.~\ref{fig:ek_growth} and Fig.~\ref{fig:mixing_width}.

The scalar diagnostics do not fully characterize the nonlinear density morphology. To provide a more comprehensive assessment, Appendix~\ref{app:2d_fourier} repeats the same experiments using a two-dimensional 90\%-power spectral reconstruction. The reconstructed perturbations reproduce the same vertically coherent, plume-like morphology and nonlinear evolution described above, further confirming the conclusions drawn from the one-dimensional analysis.

\section{Discussion and Conclusions}
\label{sec:conclusions}

This work introduces the CNOP framework as a quantitative approach for identifying optimal perturbations and assessing finite-time predictability in astrophysical fluid dynamics. In the compressible two-phased astrophysical fluid, the nonlinear evolution is highly sensitive to the initial perturbation, making the identification of the most dangerous disturbances particularly important. Unlike traditional non-normal linear analysis, which provides only qualitative insights into transient growth, the CNOP method offers a fully nonlinear and quantitative characterization of optimal perturbations.  For the compressible RT instability considered here, the most dangerous initial perturbations are found to be coherent, interface-localized, multi-mode structures whose detailed morphology depends on the optimization time horizon.

Our results demonstrate that the preferred scales of the optimized perturbation does not correspond to a classical fastest-growing RT mode. In the inviscid sharp-interface limit, the linear growth rate, $\sigma=\sqrt{Agk}$, increases monotonically with the wavenumber $k$, implying that linear theory does not predict a finite preferred wavelength. A plausible qualitative picture is that the finite interface thickness $h_s=0.02\,L$ (Eq.~\ref{eq:rho0}) suppresses the growth of sufficiently short wavelengths ($kh_s\gtrsim1$), whereas the finite optimization time limits the amplification of long wavelength disturbances. The CNOP optimization therefore selects perturbations from an intermediate range of scales. For the dominant mode, $n\sim10$, one obtains $kh_s\sim1.3$, consistent with this interpretation. The nearly identical spectra obtained at the optimization horizon $t^*=2$ on both $N=128$ and $N=256$ grids further indicate that the preferred band is a physical feature rather than a numerical artifact. When the optimization horizon is extended to $t^*=3$, the spectrum broadens toward lower wavenumbers, which suggests that the optimizer increasingly exploits longer-wavelength structures. These structures subsequently evolve into coherent large-scale bubbles and spikes during the nonlinear development. Although no analytical theory currently predicts the preferred spectral band uniquely, our results indicate that it emerges naturally from the interplay among finite interface thickness, finite optimization time, and nonlinear dynamics embodied in the finite-time CNOP optimization problem.

The astrophysical implication extend well beyond the idealized RT problem considered here. If an astrophysical system exhibits more efficient instability-driven mixing than expected from weak, incoherent perturbations, the discrepancy does not necessarily imply a larger seed energy. Instead, it may reflect a more coherent seed structure. Examples include convective plumes in a stellar envelope or progenitor \citep{Meakin2007}, large-amplitude convective fluctuations and variability in radiation-pressure-supported massive-star envelopes that are absent from one-dimensional models \citep{Jiang2018MassiveEnvelopes}, coherent corrugations of supernova ejecta and circumstellar shells \citep{Wang2001,Kifonidis2003,luo2026cocoon}, periodically modulated AGN jets and winds \citep{yuan2018active,zhang2025macer3d}, and magnetic and vortical structures that imprint phase coherence on an interface. Similar issues arise in many other astrophysical environments, including accretion disks, where MHD instabilities regulate angular momentum transport \citep{balbus1991powerful}, magnetized galactic disks subject to buoyancy and Parker-like instabilities \citep{parker1958dynamical}, and the multiphase circumgalactic medium \citep{faucher2023key}, where cloud survival and gas exchange depend on the structure of turbulent interfaces \citep{gronke2018growth,ji2019simulations,tan2021radiative}. In this broader context, CNOP provides a quantitative framework for linking the observed or simulated efficiency of instability-driven mixing to the coherence and structure of otherwise uncertain initial perturbations, thereby offering a new approach for constraining seed mechanisms across a wide range of astrophysical fluid systems.

Several limitations should be noted. The present calculations are two-dimensional, inviscid compressible Euler simulations with an ideal-gas equation of state, without explicit cooling and magnetic field, and the perturbations restricted to the vertical velocity component. The optimization targets a single objective, the squared kinetic-energy density anomaly, evaluated at one prescribed time. Other astrophysical applications may require objective functions based on mixed mass, vorticity, magnetic energy, emissivity, or tracer transport. Finally, because the objective for CNOP is inherently nonconvex,  convergence to the similar solutions from multiple random initializations provides encouraging evidence of robustness  but not aconstitute  mathematical proof of a global optimum. Nevertheless, the present results establish CNOP as a viable framework for astrophysical fluid dynamics. Unlike conventional linear analyses, CNOP directly identifies the initial perturbations that drive the strongest nonlinear evolution, thereby providing a quantitative framework for investigating instability, transition, and predictability in complex astrophysical systems. The approach is readily extendable to magnetized, radiative, and multiphase flows, potentially enabling a systematic exploration of nonlinear pathways that shape the evolution of astrophysical environments across a wide range of scales.

Our main conclusions are:
\begin{enumerate}
  \item \emph{CNOP formulation for compressible astrophysical flows} CNOP can be formulated and solved for a compressible astrophysical hydrodynamic instability using differentiable finite-volume simulations, providing a quantitative framework for identifying finite-time nonlinear optimal perturbations. 
  \item \emph{Structure and spectral characteristics of optimal seeds} At a fixed perturbation energy of $\varepsilon=10^{-6}$, the optimized seeds are coherent, interface-localized, and multimodal. At $t^*=2$, the dominant bands are $n=9$--$15$ at the grid size $N = 128$ and $n=9$--$16$ at the grid size $N = 256$. Their near resolution independence indicates that the preferred scales are physical rather than grid-scale artifacts. For $t^*=3$,  the spectrum becomes broader and more dispersed. 
  \item \emph{Enhanced nonlinear instability growth} Compared with random perturbations of the same energy, the CNOP seeds produce substantially stronger nonlinear RT evolution, increasing the kinetic energy by factors of $20.7$ and $32.5$ in kinetic energy at $t^*=2$ for the grid sizes, $N=128$ and $N=256$, respectively, and by a factor of $8.6$ at $t^*=3$;
  \item \emph{Dominant-mode representation of nonlinear evolution} A reconstruction from the five dominant Fourier modes captures much of the nonlinear growth and mixing at $t^*=2$. The larger deviations in kinetic energy and morphology at $t^*=3$ indicate that weaker, more broadly distributed spectral components become increasingly important during extended nonlinear evolution.
  \item \emph{Mixing enhancement as an emergent consequence} Although mixing is not included in the optimization objective, the optimized seeds generate substantially broader mixing layers. CNOP therefore enhances physical mixing as an emergent consequence of optimized kinetic-energy amplification rather than merely increasing velocity fluctuations. 
  \item \emph{Finite-time nonlinear predictability and preferred scales} The optimal perturbation structure depends on the target time, demonstrating that finite-time nonlinear predictability contains information beyond classical linear normal-mode analysis. The preferred wavelength results from the competition between rapid RT amplification and suppression of short wavelengths by the finite interface thickness. For longer optimization horizons, the spectrum shifts toward larger scales as the optimizer increasingly exploits nonlinear bubble-and-spike dynamics. These results suggest that instability-driven mixing in astrophysical systems depends not only on the amplitude of seed perturbations but also on their spatial organization and coherence, which are shaped by the underlying flow and interface structure. 
\end{enumerate}

\begin{acknowledgments}
    We thank Y.S.Ting for helpful discussions. The authors are supported by the Natural Science Foundation of China (grants 12522301, 12241105, 12192223 and 12361161601), SIMIS-ID-2025-CL, the National Key R\&D Program of China (No. 2023YFB3002502), and the China Manned Space Program (grants CMS-CSST-2025-A08 and CMS-CSST-2025-A10). Numerical calculations were run on the high-performance computing resource on the CFFF platform of Fudan University and in the Supercomputing Center of Wuhan University.
\end{acknowledgments}

\software{{\small Matplotlib} \citep{hunter2007matplotlib},
          {\small NumPy} \citep{harris2020array},
          {\small SciPy} \citep{virtanen2020scipy},
          {\small yt} \citep{turk2010yt,turk2025introducing},
          {\small Astronomix} \citep{storcks_astronomix_2025},
          {\small JAX} \citep{Bradbury2018}
          }

\restartappendixnumbering
\appendix

\section{Convergence of the CNOP Optimizer}
\label{app:convergence}

\begin{figure*}[!ht]
  \centering
  \includegraphics[width=\textwidth]{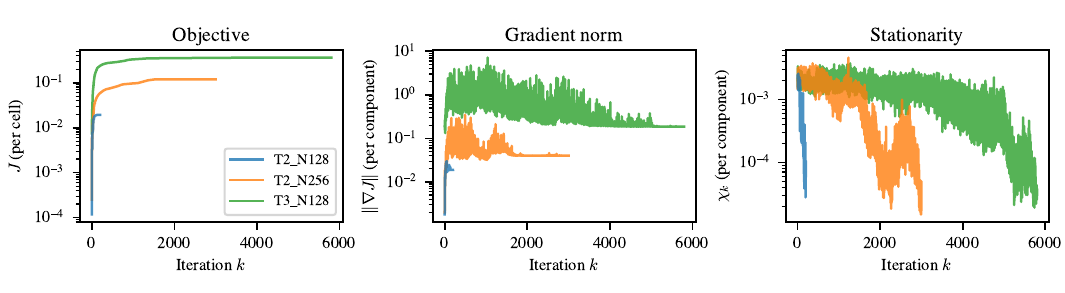}
  \caption{Convergence history of the objective magnitude (left), gradient norm (center), and stationarity measure $\chi_k$ (right) versus iteration number for {\tt T2\_N128}, {\tt T2\_N256}, and {\tt T3\_N128}.}
  \label{fig:convergence_history}
\end{figure*}

The stationarity measure is defined as $\chi_k = \|P(\delta u_k - \nabla J_k) - \delta u_k\|_\infty$, where $P$ is the projection operator defined in Eq.~\eqref{eq:projection}. A small value of $\chi_k$ indicates that the current iterate lies near a stationary point of the projected Lagrangian, providing a resolution-independent convergence criterion. All three runs ({\tt T2\_N128}, {\tt T2\_N256}, and {\tt T3\_N128}) achieve satisfactory convergence within the iteration budgets listed in Tab.~\ref{tab:runs}, as evidenced by the plateau in $-J$ and the decay of $\chi_k$ shown in Fig.~\ref{fig:convergence_history}.

\section{Two-dimensional Fourier Reconstruction}
\label{app:2d_fourier}

\begin{figure*}[!ht]
  \centering
  \includegraphics[width=0.8\textwidth]{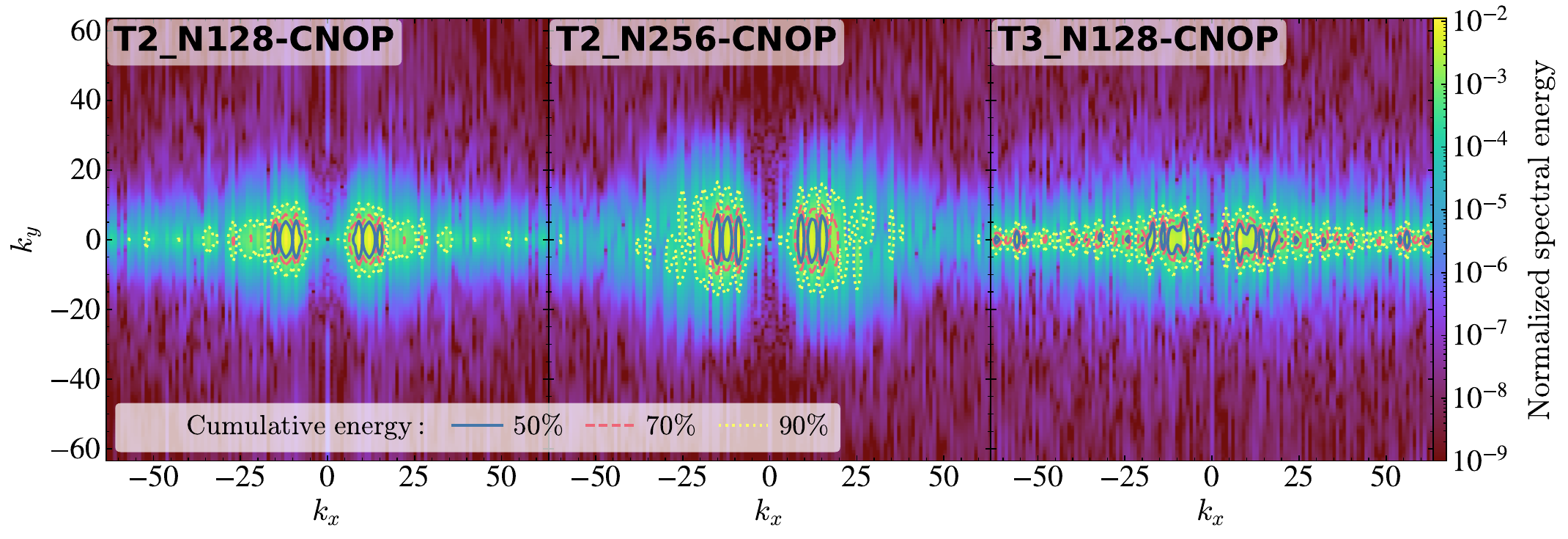}
  \caption{Normalized spectral energy $E_{n_x, n_y}$ of CNOPs across the two-dimensional Fourier modes $(n_x, n_y)$. The contours enclose 50\%, 70\%, and 90\% of the cumulative spectral energy.}
  \label{fig:fourier_2d}
\end{figure*}

The five-mode horizontal reconstruction utilized in the main text serves as a deliberately restrictive test to isolate the contribution of the dominant horizontal Fourier band. To provide a more comprehensive perspective, we present the two-dimensional spectral-energy distributions of the CNOPs in Fig.~\ref{fig:fourier_2d}. 
By examining the contours enclosing 50\%, 70\%, and 90\% of the total Fourier spectral energy, we observe that the spectral energy remains concentrated in compact regions of Fourier space. This localization supports the use of the one-dimensional horizontal spectral decomposition in~\S~\ref{sec:fourier_structure}, which effectively captures the dominant features of the CNOP perturbation structure. Specifically, as shown in the left panel of Fig.~\ref{fig:fourier_2d}, the baseline case {\tt T2\_N128} exhibits this concentrated spectral distribution, with a slightly broader extent of the high-energy region in the vertical wavenumber direction. In the middle panel of Fig.~\ref{fig:fourier_2d}, corresponding to the higher-resolution case {\tt T2\_N256}, the horizontal spectral concentration remains consistent with the one-dimensional analysis in~\S~\ref{sec:fourier_structure}. However, the two-dimensional spectrum reveals a slight spreading in the vertical wavenumber direction as the resolution increases, a feature that cannot be captured by the one-dimensional horizontal spectral decomposition. In the right panel of Fig.~\ref{fig:fourier_2d}, corresponding to the longer-time horizon case {\tt T3\_N128}, the horizontal spectral distribution becomes more dispersed, again consistent with the one-dimensional analysis in~\S~\ref{sec:fourier_structure}. Meanwhile, the vertical spectral extent shows no substantial increase, indicating that the dominant change at later times occurs primarily in the horizontal spectral distribution. 

\begin{figure*}[!ht]
  \centering
  \includegraphics[width=\textwidth]{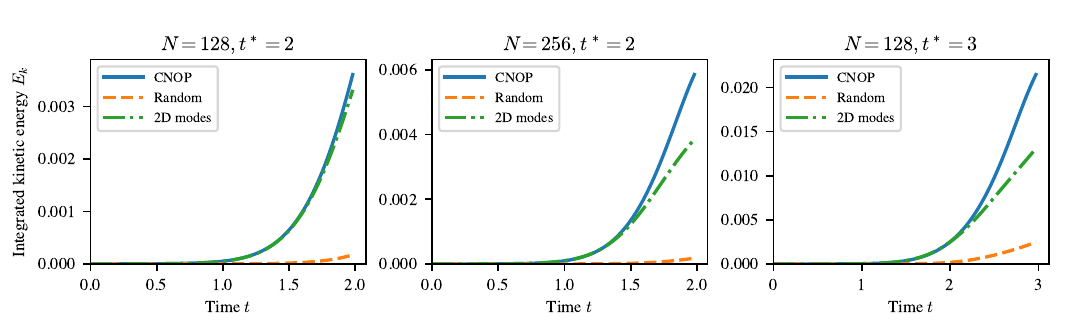}
  \caption{The comparison of total kinetic energy growth generated by CNOPs.}
  \label{fig:ek_2d}
\end{figure*}

\begin{figure*}[!ht]
  \centering
  \includegraphics[width=\textwidth]{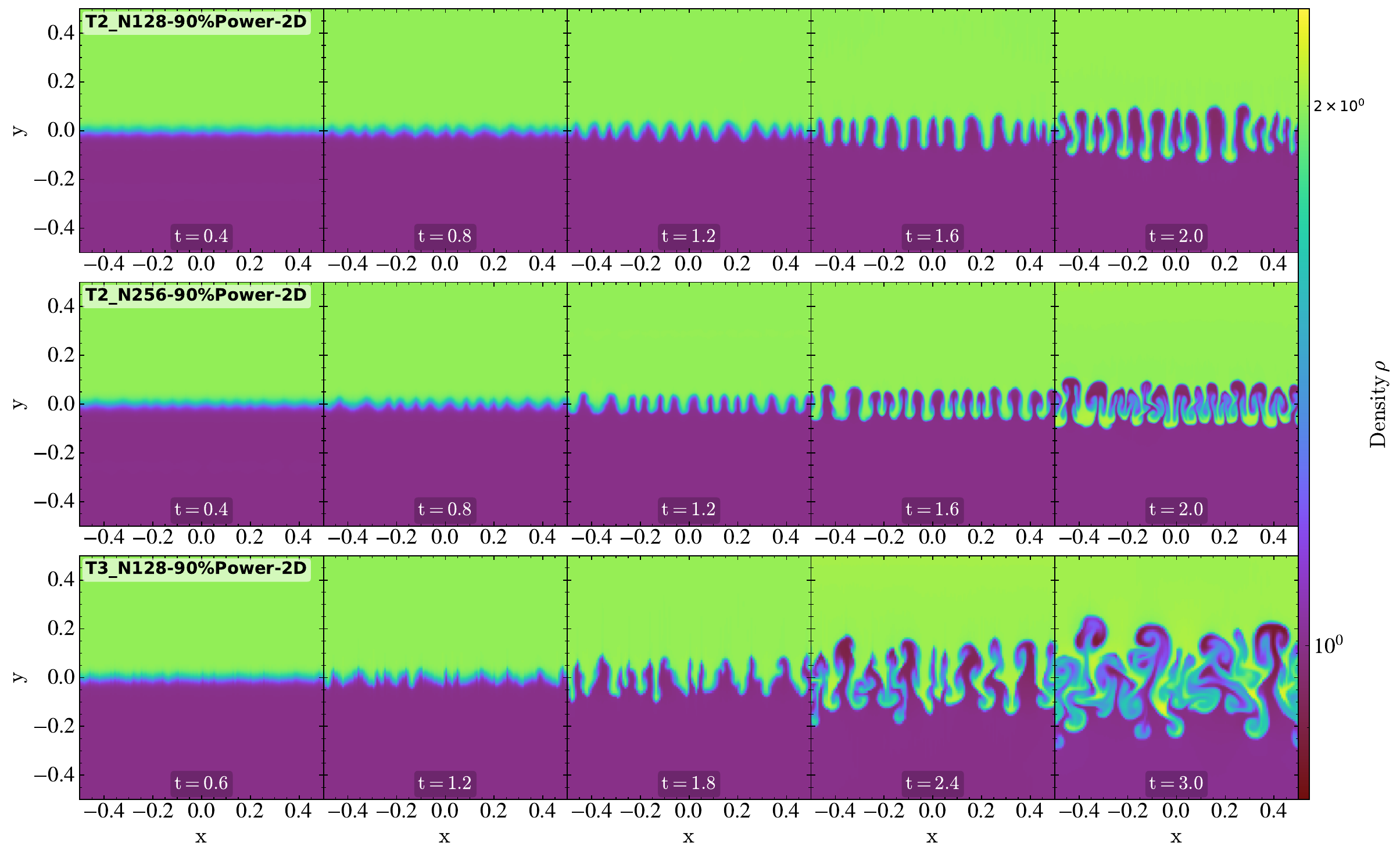}
  \caption{Nonlinear density evolution for the two-dimensional 90\% spectral-energy Fourier reconstructions. Rows (top to bottom) show {\tt T2\_N128-90\%Power-2D}, {\tt T2\_N256-90\%Power-2D}, and {\tt T3\_N128-90\%Power-2D}.}
  \label{fig:density_evolution_2d}
\end{figure*}

Figs.~\ref{fig:ek_2d} and~\ref{fig:density_evolution_2d} show that a two-dimensional Fourier reconstruction retaining 90\% of the CNOP spectral energy captures both the rapid kinetic-energy growth and the coherent plume-like density evolution. In particular, a direct comparison of Fig.~\ref{fig:density_evolution_2d} with the full-CNOP and five-dominant-mode evolutions in Fig.~\ref{fig:density_evo} shows that retaining the vertical spectral content and the associated two-dimensional phase relationships better restores the organized bubble-and-spike development. However, as the time horizon increases, the reconstruction (bottom panel of Fig.~\ref{fig:density_evolution_2d}) departs more visibly from the full-CNOP evolution (Fig.~\ref{fig:sim_evo_t3}). This divergence highlights the increasing complexity and the cascade of nonlinear interactions that occur over longer durations, which cannot be adequately reconstructed from the initial spectral components alone.

\bibliography{ref}{}
\bibliographystyle{aasjournalv7}

\end{document}